\documentclass[
aps,
twocolumn,
superscriptaddress,
floatfix,
longbibliography,
nofootinbib
]{revtex4-1}
\usepackage[dvipdfmx]{graphicx}
\usepackage{dcolumn}
\usepackage{bm}
\usepackage{ulem} 
\usepackage{amsmath}
\usepackage{amssymb}
\usepackage{txfonts}
\usepackage{hyperref}
\usepackage{color} 
\usepackage{xcolor}
\usepackage{listings}
\usepackage{cprotect}

\hypersetup{
  colorlinks=true,
  linkcolor=[rgb]{0.00,0.00,1.00},
  citecolor=[rgb]{0.00,0.00,1.00},
  urlcolor=[rgb]{0.00,0.00,1.00},
  setpagesize=false
}

\begin{document}
\title{Optical absorption activated by an ultrashort half-cycle pulse in metallic and
superconducting states of the Hubbard model}
\author{Kazuya Shinjo}
\affiliation{Computational Quantum Matter Research Team, RIKEN Center for Emergent Matter Science (CEMS), Wako, Saitama 351-0198, Japan}
\author{Shigetoshi Sota}
\affiliation{Computational Materials Science Research Team,
RIKEN Center for Computational Science (R-CCS), Kobe, Hyogo 650-0047, Japan}
\affiliation{Quantum Computational Science Research Team,
RIKEN Center for Quantum Computing (RQC), Wako, Saitama 351-0198, Japan}
\author{Seiji Yunoki}
\affiliation{Computational Quantum Matter Research Team, RIKEN Center for Emergent Matter Science (CEMS), Wako, Saitama 351-0198, Japan}
\affiliation{Computational Materials Science Research Team,
RIKEN Center for Computational Science (R-CCS), Kobe, Hyogo 650-0047, Japan}
\affiliation{Quantum Computational Science Research Team,
RIKEN Center for Quantum Computing (RQC), Wako, Saitama 351-0198, Japan}
\affiliation{Computational Condensed Matter Physics Laboratory,
RIKEN Cluster for Pioneering Research (CPR), Saitama 351-0198, Japan}
\author{Takami Tohyama}
\affiliation{Department of Applied Physics, Tokyo University of Science, Tokyo 125-8585, Japan}

\date{\today}
             
 
\begin{abstract}
The development of high-intensity ultrashort laser pulses unlocks the potential of pump-probe spectroscopy in sub-femtosecond timescale. 
Notably, subcycle pump pulses can generate electronic states unreachable by conventional multicycle pulses, leading to a phenomenon that we refer to as subcycle-pulse engineering. 
In this study, we employ the time-dependent density-matrix renormalization group method to unveil the transient absorption spectra of superconducting and metallic states in nearly half-filled one-dimensional and two-dimensional Hubbard models excited by an ultrashort half-cycle pulse, which can induce a current with inversion-symmetry breaking. 
In a superconducting state realized in attractive on-site interactions, we find the transient activation of absorptions at energies corresponding to the amplitude modes of superconducting and charge-density-wave states.
On the other hand, in a metallic state realized in the two-dimensional model with repulsive on-site interactions, we obtain another type of absorption enhancements, which are distributed broadly in spin excitation energies.
These findings indicate that superconducting and metallic states are sensitive to an ultrashort half-cycle pulse, leading to the transient activations of optical absorptions with their respective mechanisms.
\end{abstract}
\maketitle

%
\section{Introduction}\label{sec1}
The development of strong ultrashort pulse lasers has enabled the observation of non-equilibrium electron dynamics on femtosecond and even attosecond timescales~\cite{Krausz2009, Gallmann2012}.
A novel technique called subcycle-pulse engineering has been proposed, involving the application of a subcycle pulse with less than one cycle oscillation within a pulse envelope~\cite{Tsuji2012, Lenarcic2014, Shinjo2022, Shinjo2023, Tohyama2023, Shinjo2024}.
This approach aims to generate electronic states that cannot be accessed using conventional multicycle pulses.
It has been suggested that applying ultrashort subcycle pulses to metallic or superconducting states can induce current with inversion-symmetry breaking, which is detectable by second-harmonic generations~\cite{Kawakami2020, Shinjo2023}.

In a current-induced state with a broken inversion symmetry, exotic phenomena are anticipated, including the Doppler shifts of quasi-particle excitations known as the Volovik effect~\cite{Volovik1993}. 
Particularly in absorption spectra, supercurrents in a dirty superconductor with intermediate strength of disorder can activate absorptions at a Higgs-mode energy $\omega=2\Delta_\text{s}$ between quasiparticle excitations with a gap $\Delta_\text{s}$~\cite{Moor2017}.
This behavior has been observed in a superconductor NbN with a current generated by a DC bias~\cite{Nakamura2019, Shimano2020}. 
While disorder effects are often essential in understanding absorptions in superconductors~\cite{Mattis1958, Laplae1983, Zimmermann1991}, it has been pointed out that inversion-symmetry breaking due to supercurrent can change a selection rule, leading to a significant absorption at $\omega=2\Delta_\text{s}$ even in clean superconductors~\cite{Seibold2020,Papaj2022, Crowley2022}.

In this paper, we investigate how ultrashort half-cycle pulses, which can induce a current with an inversion-symmetry breaking, contribute to transient absorption spectra of strongly correlated electron systems.
Using the time-dependent density-matrix renormalization group (tDMRG) method, we demonstrate that the half-cycle pump pulse leads to the enhancement of optical absorptions in metallic and superconducting states of the Hubbard model.

In the one-dimensional (1D) and two-dimensional (2D) Hubbard models with attractive $U<0$ on-site interactions, we find the enhancement of absorptions at energies corresponding to amplitude modes associated with superconducting and charge-density-wave (CDW) orders.
This behavior is the manifestation of a current-induced activation of absorptions at amplitude modes in a clean superconducting state.
In a 2D system, we find that the amplitude-mode absorptions emerge in both parallel and perpendicular directions to the pump pulse for $U=-4$ (in the unit of the electron hopping), but only in parallel for $U=-8$.
This $U$ dependence comes from the difference in the localized nature of superconducting pairing at weak and strong couplings.

In a metallic state of the 2D Hubbard model at nearly half filling with repulsive $U>0$ on-site interactions, we find another type of absorption enhancement broadly distributed in the Mott gap.
The mid-gap absorptions appear in both directions parallel and perpendicular to the pump pulse, whose nonlocal nature can be attributed to the effect of magnetic excitations.

The rest of this paper is organized as follows.
We introduce the Hubbard model with a time-dependent electric field in Sec.~\ref{sec:2}.
We also show a method to obtain time-dependent optical conductivity by the tDMRG method.
In Sec.~\ref{sec:3}, we show transient absorption spectra of the Hubbard models with attractive interactions $U<0$.
We demonstrate that a half-cycle pulse activates absorptions at superconducting and CDW amplitude modes.
In Sec.~\ref{sec:4}, we show transient absorption spectra of the Hubbard models with repulsive interactions $U>0$.
We show the enhancement of absorptions at energies corresponding to spin-exchange interactions.
Finally in Sec.~\ref{sec:5}, we give a summary of this paper.
Bond dimension dependence is examined in Appendix~\ref{sec:A1} and additional results for the $t$-$J$ model are provided in Appendix~\ref{sec:A2}.

%
\section{Model and Method}\label{sec:2}
To investigate nonequilibrium properties of strongly correlated electron systems, we consider the Hubbard model on a ladder-shaped cluster of $L=L_{x}\times L_{y}$ sites with a time ($t$) dependent vector potential 
$\bm{A}(t)=[A_{x}(t),A_{y}(t)] = A_{x}(t) \bm{e}_{x} + A_{y}(t)\bm{e}_{y}$, 
with $\bm{e}_{x}$ ($\bm{e}_{y}$) being the unit vector along the $x$-axis ($y$-axis) direction, described by the following Hamiltonian:
\begin{align}\label{eq-H}
\mathcal{H}=&-t_\text{h}\sum_{\langle i,j\rangle,\sigma} B_{ij,\sigma}
+U\sum_{i}n_{i,\uparrow}n_{i,\downarrow},
\end{align}
where $B_{ij,\sigma} = e^{-i\bm{A}(t)\cdot \bm{R}_{ij}}c_{i,\sigma}^{\dag}c_{j,\sigma} + \text{H.c.}$, $c_{i,\sigma}^\dag$ is the electron creation operator at site $i$, located at $\bm{r}_{i}$ in the cluster, with spin $\sigma \,(=\uparrow,\downarrow)$, and $n_{i,\sigma}=c_{i,\sigma}^\dag c_{i,\sigma}$.
$t_\text{h}$ is nearest-neighbor hopping and $U$ is on-site Coulomb interaction. 
Hereafter, we set $t_\text{h}=1$ as the energy unit.
The sum indicated by $\langle i,j  \rangle$ runs over all pair of nearest-neighbor sites $i$ and $j$, and $\bm{R}_{ij} = \bm{r}_{i} - \bm{r}_{j}$.
We impose open boundary conditions along the $x$- and $y$-axis directions.
A spatially homogeneous electric field $\bm{E}(t) = -\partial_{t}\bm{A}(t)$ is incorporated via the Peierls substitution in the hopping terms~\cite{Peierls1933}.
Note that we set the light velocity $c$, the elementary charge $e$, the Dirac constant $\hbar$, and the lattice constant to 1.

The effect of a strong ultrashort pump pulse applied to the Hubbard model can be investigated with a vector potential $\bm{A}_\text{pump}(t)=A_\text{pump}(t)\bm{e}_{x}$ introducing a phase shift~\cite{Lenarcic2014} at $t=0$ in the $x$-axis direction with
\begin{equation}\label{eq-pump}
A_\text{pump}(t)=
\left\{ \,
    \begin{aligned}
    & 0 \;\;\;\; (t<0), \\
    & \frac{A_{0}}{t_\text{d}}t \; (0\le t\le t_\text{d}) \\
    & A_{0} \; (t>t_\text{d}),
    \end{aligned}
\right.,
\end{equation}
where we set the width of the pump pulse to $t_\text{d}=0.02$~\cite{note_td}.
For $t_\text{d} \ll t_\text{h}^{-1}$, $\bm{E}_\text{pump}(t) = -\partial_{t}\bm{A}_\text{pump}(t)$ leads to a half-cycle pulse that can quench the flux of $A_{0}$ and twists the phase of many-body wavefunctions~\cite{Shinjo2023}.
As a result, the pump pulse can induce an inversion-symmetry breaking associated with a current in metallic and superconducting states.

To obtain transient absorptions of the Hubbard model parallel (perpendicular) to the pump pulse, we subsequently apply a probe pulse $\bm{A}_\text{probe}(t,\tau)=A_\text{probe}(t,\tau)\bm{e}_{x}$ [$A_\text{probe}(t,\tau)\bm{e}_{y}$] at time $t=\tau$ polarized in the $x$-axis ($y$-axis) direction with
\begin{align}
A_\text{probe}(t,\tau)=A_0^\text{pr} e^{-\left(t-\tau \right)^2/\left[2(t_\mathrm{d}^\text{pr})^2\right]} \cos \left[\Omega^\text{pr}(t-\tau)\right].
\end{align}
We set $A_0^\text{pr}=0.001$, $\Omega^\text{pr}=10$, and $t_\text{d}^\text{pr}=0.02$.
Note that $\tau$ indicates the delay time between pump and probe pulses.
We obtain a current 
\begin{equation}
\bm{j}(t)=-\langle \partial \mathcal{H}[\bm{A}(t)]/\partial{\bm{A}(t)} \rangle 
\label{eq:j}
\end{equation}
with a time-dependent many-body wave function $|\psi (t)\rangle$ evaluated by the tDMRG method implemented with the Legendre polynomial~\cite{Shinjo2021, Shinjo2021b, Shinjo2022}. 
Here, $\langle\cdots\rangle$ implies $\langle\psi(t)|\cdots|\psi(t)\rangle$ and the initial state of $|\psi(t)\rangle$ is the ground state of the Hubbard model $\mathcal{H}$ with $\bm{A}(t)=0$ in Eq.~(\ref{eq-H}). 
With the present methodology, we obtain both singular and regular parts of the optical conductivity in nonequilibrium~\cite{Shao2016, Shinjo2018, Shinjo2022} by directly calculating
\begin{align}
\bm{\sigma}(\omega,\tau) = \frac{\bm{j}_\text{probe}(\omega,\tau) }{i(\omega +i\gamma)L\bm{A}_\text{probe}(\omega,\tau)},
\label{eq:sigma}
\end{align}
where $\bm{A}_{\text{probe}}(\omega,\tau)$ and $\bm{j}_\text{probe}(\omega,\tau)$ are the Fourier transform of $\bm{A}_\text{probe}(t,\tau)$ and $\bm{j}_\text{probe}(t,\tau) = \bm{j}_\text{total}(t,\tau)-\bm{j}_\text{pump}(t)$, respectively.
Here, $\bm{j}_\text{total}(t,\tau)$ and $\bm{j}_\text{pump}(t)$ are obtained by taking $\bm{A}(t)=\bm{A}_\text{pump}(t)+\bm{A}_\text{probe}(t,\tau)$ and $\bm{A}(t)=\bm{A}_\text{pump}(t)$, respectively, in Eq.~(\ref{eq:j}).
For the Fourier transformation, we evaluate $\bm{j}_\text{probe}(t,\tau)$ for $t$ up to $t_\text{max}=20$.
We take a broadening factor $\gamma=0.4>2\pi/t_\text{max}$ in this paper.
In a finite system with open boundary conditions, we obtain the singular part of $\sigma^{xx (yy)}(\omega,\tau)$ at $\omega \propto 1/L>0$, where $\sigma^{xx}(\omega,\tau)$ and $\sigma^{yy}(\omega,\tau)$ are the diagonal components of $\bm{\sigma}(\omega,\tau)$ along the $x$- and $y$-axis directions, respectively.

%
\section{Attractive Hubbard model}\label{sec:3}
\subsection{One dimensional systems}

\begin{figure}[t]
\includegraphics[width=0.45\textwidth]{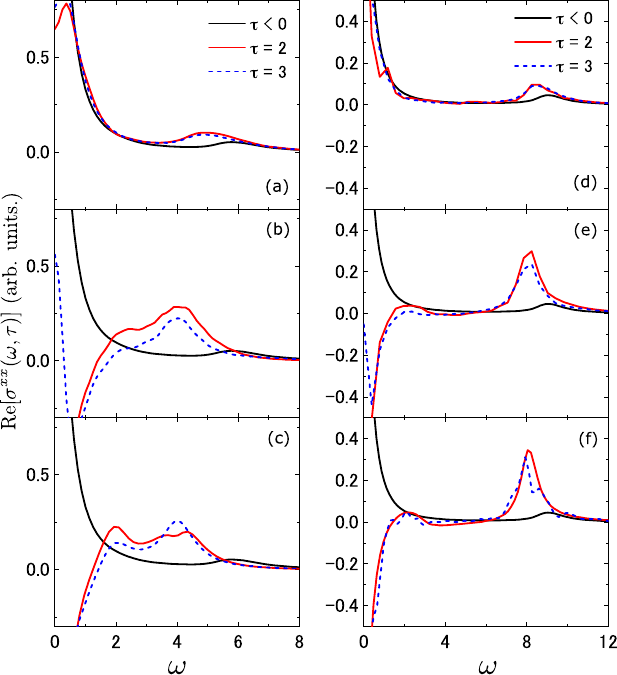}
\caption{Re~$[\sigma^{xx}(\omega,\tau)]$ of the $\delta=1/8$ Hubbard model on a 1D cluster with $L=32\times1$ 
for (a)--(c) $U=-4$ and (d)--(f) $U=-8$. 
The black lines are for $\tau<0$ (i.e., before pumping) and the red solid and blue dashed lines are for $\tau=2$ and 3, respectively, 
with pumping amplitudes $A_{0}=0.1\pi$ [(a) and (d)], $A_{0}=0.4\pi$ [(b) and (e)], and $A_{0}=0.5\pi$ [(c) and (f)].
}
\label{Fig1}
\end{figure}

We first show transient absorption spectra $\text{Re}[\sigma^{xx}(\omega,\tau)]$ for the attractive Hubbard model with $U<0$ on a 1D cluster $L=32\times1$.
We set $U=-4$ in Figs.~\ref{Fig1}(a)--\ref{Fig1}(c) and $U=-8$ in Figs.~\ref{Fig1}(d)--\ref{Fig1}(f), which respectively lie on the Bardeen-Cooper-Schrieffer (BCS) and Bose-Einstein-condensation (BEC) sides across the BCS-BEC crossover~\cite{Scalettar1989, Moreo1991, Fontenele2022, Hartke2023}.
These results are obtained by doping electrons with density $\delta = N/L -1 = 1/8$ in a half-filled state, where $N$ is the total number of electrons.
However, note that since we consider bipartite lattices in this study, a particle-hole transformation maps electron doping to hole doping.

Before applying a pump pulse, i.e., $\tau<0$, we find that most of spectral weights concentrate at $\omega \simeq 0$ due to the presence of a nonzero superfluid weight $\sigma_\text{S}> 0$, as shown in the black lines in Fig.~\ref{Fig1}.
In other words, most of contributions to Re$[\sigma^{xx}(\omega,\tau<0)]$ comes from the diamagnetic term $-T_{x}(\tau<0)=\frac{t_\text{h}}{L}\sum_{\langle i,j\rangle,\sigma} \langle \psi(\tau)| B_{ij,\sigma} |\psi(\tau)\rangle$.
Here, note that the peak structure has a finite width since a broadening factor $\gamma$ is introduced in Eq.~(\ref{eq:sigma}).

For delay times $\tau=2$ and 3 after pumping, indicated by red solid and blue dashed lines in Fig.~\ref{Fig1}, respectively, we find the suppression of $\sigma_\text{S}$ and the enhancement of absorptions at $\omega>0$, which become clearer with increasing $A_{0}$~\cite{notex}.
In particular at $U=-4$, we find the enhancement of a lower-energy mode at $\omega \simeq2$, as shown in Figs.~\ref{Fig1}(b) and \ref{Fig1}(c).
We consider that this structure corresponds to the amplitude mode of a superconducting order, i.e., the Higgs mode activated by the inversion-symmetry breaking induced by a half-cycle pulse~\cite{notesc}.
The breaking of an inversion symmetry via a current in a superconducting state results in an absorption at $\omega=2\Delta_\text{s}$ in the optical conductivity, where $\Delta_{s}$ is a superconducting gap~\cite{Papaj2022, Crowley2022}.
For $U=-4$, we can roughly estimate $\Delta_\text{s}=1.4$ with the BCS mean-field model for half filling on a square lattice.
Considering the suppression of $\Delta_{s}$ by quantum fluctuations~\cite{Gersch2008} to about 70\%, it is reasonable for an activated Higgs mode to appear at $\omega=2$.
The Higgs-mode absorptions observed in the 1D system do not appear stable for long times and are captured by ultrafast transient absorption spectroscopy.

At strong coupling $U=-8$, there is no enhancement in absorption associated with the Higgs mode. 
This is likely associated with the fact that the Higgs mode becomes broadened and eventually disappears when the Cooper pairs turn into tightly bound pairs leading to BEC~\cite{Varma2002, Pekker2015, Behrle2018}.

In addition to the Higgs-mode absorptions, we find another enhancement in absorptions at higher energies $\omega \simeq |U|$, as shown in Figs.~\ref{Fig1}(b) and \ref{Fig1}(c) [Figs.~\ref{Fig1}(e) and \ref{Fig1}(f)] for $U=-4$ ($U=-8$).
This is due to the activation of another amplitude mode, i.e., a CDW amplitude mode, by the half-cycle pulse, since $|U|$ corresponds to the energy required to dissolve a doublon.

We also find that the kinetic energy $T_{x}(t)$ increases with $A_{0}$ and almost zero at $A_{0}=0.5\pi$ for $U=-4$ and $-8$~\cite{Note1}.
This behavior indicates that the total spectral weights proportional to $-T_{x}(t)$ is also almost zero after pumping at $A_{0}=0.5\pi$, although the amplitude-mode absorptions with positive spectral weights are enhanced.
The enhanced contribution of the paramagnetic Van Vleck term with spectral weights $-T_{x}(t) - \sigma_{S}$ is allowed because
the decrease in total spectral weights is compensated by the emergence of a negative superfluid weight $\sigma_\text{S}<0$~\cite{Stafford1991, Fye1991, Fye1992}.

\subsection{Two dimensional systems}

\begin{figure}[t]
\includegraphics[width=0.45\textwidth]{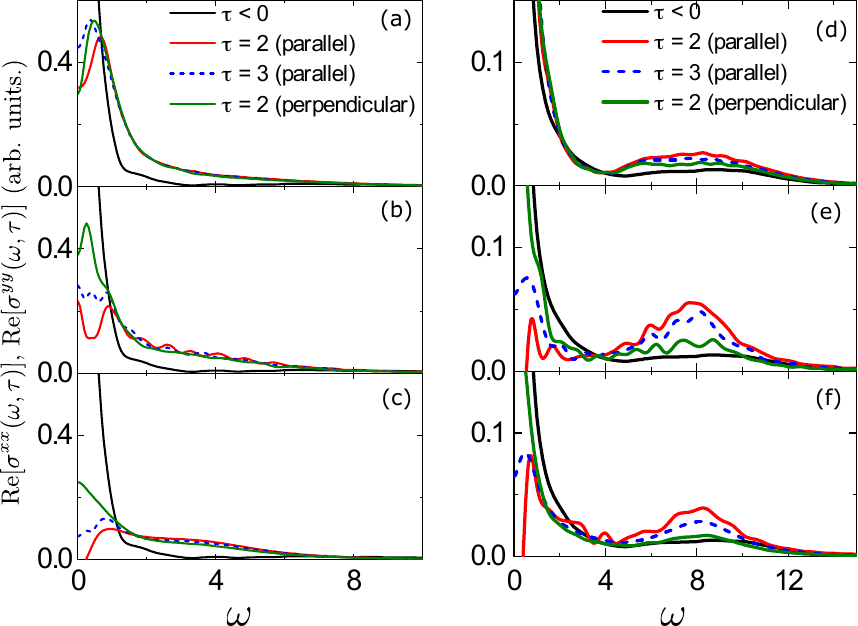}
\caption{Re$\left[ \sigma^{xx}(\omega,\tau) \right]$ (parallel to a pump pulse) and Re$\left[ \sigma^{yy}(\omega,\tau) \right]$ (perpendicular to a pump pulse) of the $\delta=1/9$ Hubbard model on a square lattice with $L=6\times 6$ for (a)--(c) $U=-4$ and (d)--(f) $U=-8$.
	The black lines show Re$\left[ \sigma^{xx}(\omega,\tau<0) \right]$ (before pumping).
	The red solid and blue dashed lines show Re$\left[ \sigma^{xx}(\omega,\tau) \right]$ at $\tau=2$ and 3, respectively, with pumping amplitudes $A_{0}=0.16\pi$ [(a) and (d)], $A_{0}=0.48\pi$ [(b) and (e)], and $A_{0}=0.64\pi$ [(c) and (f)].
	The green solid lines show Re$\left[ \sigma^{yy}(\omega,\tau) \right]$ at $\tau=2$ with $A_{0}=0.16\pi$ [(a) and (d)], $A_{0}=0.48\pi$ [(b) and (e)], and $A_{0}=0.64\pi$ [(c) and (f)].
	}
\label{Fig2}
\end{figure}

To examine the possibility of the emergence of amplitude-mode absorptions in 2D systems, we investigate Re$\left[ \sigma^{xx}(\omega,\tau) \right]$ and Re$\left[ \sigma^{yy}(\omega,\tau) \right]$ of the $\delta = 1/9$ Hubbard model $\mathcal H$ in Eq.~(\ref{eq-H}) on a square lattice with $L=6\times 6$ excited by an ultrashort half-cycle pulse.
We take the bond dimension $\chi=7000$ in the tDMRG calculations, leading to a truncation error less than $10^{-4}$.
Here, we confirm that Re$\left[ \sigma^{xx}(\omega,\tau) \right]$, obtained for a given parameter, does not change qualitatively between $\chi=6000$ and 7000 (see Appendix~\ref{sec:A1}).
We map a $6\times6$ cluster onto a 1D system using tilted-z mapping, which makes the electronic state obtained with finite $\chi$ symmetrical in the $x$- and $y$-axis directions~\cite{Shinjo2021b}. 
The results for $U=-4$ ($U=-8$) are shown in Figs.~\ref{Fig2}(a)--(c) [Figs.~\ref{Fig2}(d)--(f)] with 
$A_{0}=0.16\pi$, $0.48\pi$, and $0.64\pi$. 
Before applying a pump pulse, most of spectral weights concentrate at $\omega =0$ due to the presence of $\sigma_\text{S}>0$, as indicated by the black lines in Fig.~\ref{Fig2}.

In Fig.~\ref{Fig2}, we show Re$\left[ \sigma^{xx}(\omega,\tau) \right]$ with red solid (blue dashed) lines for $\tau=2$ ($\tau=3$) and Re$\left[ \sigma^{yy}(\omega,\tau) \right]$ with green solid line for $\tau=2$, which are probed in parallel and perpendicular to the pump pulse applied in the $x$-axis direction, respectively.
We find the enhancement of absorption at $\omega=|U|$, which corresponds to a CDW amplitude mode also observed in the 1D Hubbard model with $U=-4$ and $-8$.
In 2D systems, however, the enhanced spectral weights are broader compared to those in the 1D systems.
Thus, the Higgs-mode absorption is hidden by the CDW amplitude-mode absorption, making it difficult to distinguish between the two absorptions.
To clearly differentiate these two amplitude modes, higher resolution is required, which necessitates longer time evolutions and larger clusters in the computation. 
This exceeds our current computational resource and is therefore left for future study.

We find that the nature of the amplitude-mode absorptions varies depending on the strength of $U$.
For $U=-4$, the amplitude-mode absorptions appear in both parallel and perpendicular directions to the pump pulse.
In contrast, for $U=-8$, the amplitude-mode absorptions primarily appear in the direction parallel to the pump pulse.
This is because the pairing is strongly coupled, i.e., BEC type with a localized nature, for $U=-8$ and thus the effect of the pump pulse does not extend in the perpendicular direction.
On the other hand, for $U=-4$, the pairing is weakly coupled, i.e., BCS type with a nonlocal nature, allowing the effect of the pump pulses to extend into the perpendicular direction as well.
The nonlocal nature of pairing observed for $U=-4$ is consistent with findings in an optical lattice for $|U|\lesssim 6$~\cite{Hartke2023}.

Finally, we compare the experimental results for NbN~\cite{Nakamura2019} with our study, although a direct comparison is challenging due to differences in the experimental setup. 
In the experiment, the activation of the Higgs mode is observed only in the direction parallel to the current. 
This behavior corresponds to the amplitude-mode absorptions obtained for $U=-8$ in our study.
However, $U=-8$ is much larger than the attractive interaction present in NbN, where the superconducting state is considered to be of the BCS type.
We speculate that the disorder effect in NbN makes the nature of the amplitude-mode absorptions to be localized.

%
\section{Repulsive Hubbard model}\label{sec:4}
\begin{figure}[t]
\includegraphics[width=0.45\textwidth]{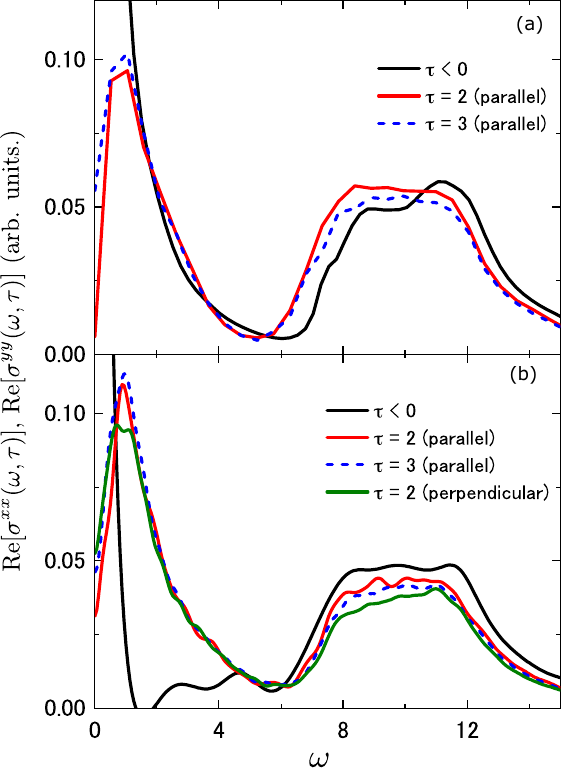}
\caption{(a) Re$[\sigma^{xx}(\omega,\tau)]$ (parallel to a pump pulse) for the repulsive Hubbard model with $U=10$ and 
$\delta=1/8$ on a $L=16\times 2$ cluster.
		(b) Re$[\sigma^{xx}(\omega,\tau)]$ (parallel to a pump pulse) and Re$[\sigma^{yy}(\omega,\tau)]$ (perpendicular to a pump pulse) for the repulsive Hubbard model with $U=10$ and $\delta=1/9$ on a $L=6\times 6$ cluster. 
		The black lines show Re$[\sigma^{xx}(\omega,\tau<0)]$ before pumping.
		The red solid and blue dashed lines represent Re$[\sigma^{xx}(\omega,\tau)]$ at $\tau=2$ and 3, respectively, 
		with a pumping amplitude of $A_{0}=0.3\pi$.   
		In (b), the green solid line represents Re$[\sigma^{yy}(\omega,\tau)]$ at $\tau=2$ with the same pump amplitude of 
		$A_{0}=0.3\pi$.}
\label{Fig4}
\end{figure}

In the 2D Hubbard models with $U=10$~\cite{noteu}, we find another type of absorption induced by a half-cycle pulse.
The results of Re$[\sigma^{xx}(\omega,\tau<0)]$ for $L=16\times 2$ are shown in Fig.~\ref{Fig4}(a), while those of Re$[\sigma^{xx}(\omega,\tau<0)]$ and Re$[\sigma^{yy}(\omega,\tau<0)]$ for $L=6\times 6$ are shown in Fig.~\ref{Fig4}(b).
In the tDMRG calculations, the bond dimensions are set to $\chi=4500$ and $7000$ for the $L=16\times 2$ and $6\times 6$ systems, respectively.
Since we consider carrier densities of $\delta=1/8$ for the $L=16\times 2$ system and $\delta=1/9$ for the $L=6\times 6$ system, both of which are near half filling, there are absorptions above the remnant Mott gap at $\omega \gtrsim 6$, as shown by the black lines in Fig.~\ref{Fig4}.
At low energies, in addition to the Drude weight at $\omega = 0$, we find incoherent spectral weights in the ranges $2< \omega < 4$ for $L=16\times 2$ and $2< \omega < 6$ for $L=6\times 6$.
These incoherent weights originate from string-type antiferromagnetic excitations present in 2D Mott insulators~\cite{Dagotto1992, Stephan1990, Inoue1990, Jaklic2000, Tohyama2005, Shinjo2021}, and have been experimentally observed in La$_{2-x}$Sr$_{x}$CuO$_{4}$~\cite{Uchida1991} and Sr$_{14-x}$Ca$_{x}$Cu$_{24}$O$_{41}$~\cite{Osafune1997} in the mid-infrared (MIR) region.
The incoherent spectral weights in the MIR band extend to higher energies, up to $\omega \lesssim 6$ in 2D systems~\cite{Tohyama2005}, compared to $\omega \lesssim 4$ in two-leg ladder systems.
Furthermore, the Drude weight is larger in the $L=16\times 2$ system than in the $L=6\times 6$ system, since the 1D nature is retained in the two-leg ladder~\cite{Osafune1997}.

After pumping with a half-cycle pulse, we find a redshift of the remnant Mott gap in the two-leg ladder system with $L=16\times 2$, as shown in Fig.~\ref{Fig4}(a), along with the suppression of the Drude weight.
We attribute this redshift to the Stark shift, as a half-cycle pulse induces polarization~\cite{Shinjo2023}.
A similar redshift behavior of the remnant Mott gap is also observed in 1D systems (not shown).
However, in contrast to quasi-1D systems, no redshift of the remnant Mott gap is observed in a square lattice, as shown in Fig.~\ref{Fig4}(b).
This is because the induced polarization after pumping is smaller in a square lattice compared to quasi-1D systems (not shown).

In the mid-gap energy range $2<\omega<3.5$, we find a slight enhancement in absorption.
The mid-gap absorptions are attributed to magnetic excitations activated by a half-cycle pulse, with peak positions characterized by spin-exchange interactions (see Appendix~\ref{sec:A2}).
This behavior suggests that antiferromagnetic order partially remains in the nearly half-filled Mott insulator, leading to spin excitations caused by charge motion.
A similar mid-gap structure has been observed in pump-probe spectroscopy of  Sr$_{14-x}$Ca$_{x}$Cu$_{24}$O$_{41}$ using a near-infrared laser~\cite{Fukaya2015, Hashimoto2016, Shao2019}.

The mid-gap absorptions are more pronounced in the 2D system with $L=6\times 6$, as shown by colored lines in Fig.~\ref{Fig4}(b), where they are broadly distributed in the range $1 \lesssim \omega \lesssim 5$.
A similar large and broad enhancement of mid-gap absorptions has also been observed in a carrier-doped Bi-based cuprate~\cite{Peli2017}, excited by a near-infrared pulse.
If we assume that the enhanced mid-gap absorption observed in this experiment shares the same origin as our results, we expect that the structure found in Fig.~\ref{Fig4}(b) can be reproduced not only by pumping with a half-cycle pulse but also with a near-infrared multicycle pulse, as investigated in Ref.~\cite{Shinjo2017}.

Finally, we compare Re$[\sigma^{xx}(\omega,\tau=2)]$ parallel to the pump pulse [the red line in Fig.~\ref{Fig4}(b)] with Re$[\sigma^{yy}(\omega,\tau=2)]$ perpendicular to the pump pulse [the green line in Fig.~\ref{Fig4}(b)].
We find that the effect of the pump pulse extends to the perpendicular direction, causing changes in absorption both parallel and perpendicular to the pump pulse.
This behavior can be attributed to spin-charge coupling in the square lattice.
Indeed, it has been reported that charge dynamics in the direction parallel to the electric field is coupled with that in the perpendicular direction via magnetic excitations~\cite{Tsutsui2021}.

\section{Summary}\label{sec:5}
Using the tDMRG method, we investigated the transient absorption spectra of superconducting and metallic states in the Hubbard models excited by an ultrashort half-cycle pulse.
In the superconducting state of the Hubbard model with attractive $U<0$ interactions, we found an enhancement of absorption induced by the half-cycle pump pulse.
The energies of these induced absorptions correspond to the amplitude modes of superconducting and CDW orders, indicating the current-induced activation of amplitude-mode absorptions in a clean superconducting state.
In the square lattice, we found that the behavior of amplitude-mode absorptions depends on the strength of $U$. 
At $U=-8$ in the strong coupling region, where pairing is formed locally, the amplitude-mode absorptions are activated only in the direction parallel to the pump pulse.
In contrast, at $U=-4$, where pairing is formed nonlocally, absorptions are activated both in the parallel and perpendicular directions.

In the metallic state of the Hubbard model on a square lattice near half filling with repulsive $U=10$ interactions, we found a different behavior in the transient absorption spectra. 
A half-cycle pump pulse induces absorptions that are broadly distributed in the mid-gap region.
These mid-gap absorptions appear in both directions parallel and perpendicular to the pump pulse, and their nonlocal nature can be attributed to the effect of magnetic excitations.

\begin{acknowledgments}
This work was supported by Grant-in-Aid for Scientific Research (B) (Nos.~19H05825, 20H01849, 21H03455, 24K00560, and 24K02948) and 
Grant-in-Aid for Early-Career Scientists (No.~23K13066) from Ministry of Education, Culture, Sports, Science, and 
Technology (MEXT), Japan, and by JST PRESTO (Grant No. JPMJPR2013).
This work was also supported in part by the COE research grant in computational science from Hyogo Prefecture and Kobe 
City through Foundation for Computational Science. 
Numerical calculation was carried out using the HOKUSAI supercomputer at RIKEN, 
computational resources of the facilities of the Supercomputer Center at Institute for Solid 
State Physics, the University of Tokyo, and the supercomputer system at the information initiative center, Hokkaido University through 
the HPCI System Research Project (Project IDs: hp230066 and hp240038).
\end{acknowledgments}

\appendix

%
\section{Bond dimension dependence}\label{sec:A1}

\begin{figure}[b]
\includegraphics[width=0.45\textwidth]{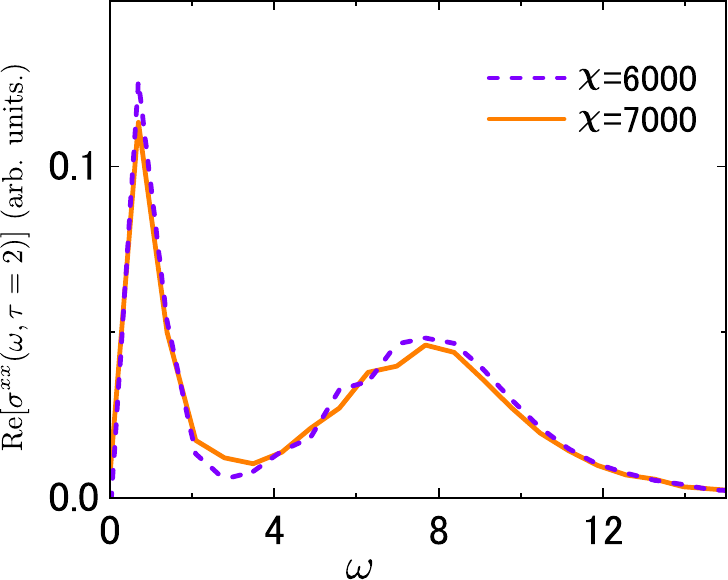}
\caption{
Re$[\sigma^{xx}(\omega,\tau=2)]$ (parallel to a pump pulse) of the $\delta=1/9$ Hubbard model on a square lattice with $L=6\times 6$ for $U=-8$, excited by a pump pulse with $A_{0}=0.3\pi$.
The purple dashed and orange solid lines represent the results obtained with $\chi=6000$ and 7000, respectively.
}
\label{FigA2}
\end{figure}

In this section, we examine bond-dimension dependency of transient absorption spectrum.
Figure~\ref{FigA2} shows Re$[\sigma^{xx}(\omega,\tau=2)]$ for the $\delta=1/9$ Hubbard model on a square lattice with $L=6\times 6$ at $U=-8$, excited by a pump pulse with $A_{0}=0.3\pi$.
Here, we demonstrate that the transient absorption spectrum remains qualitatively unchanged when comparing results with $\chi=6000$ and $\chi=7000$.
The latter value, $\chi=7000$, is specifically used for generating Figs.~\ref{Fig2} and \ref{Fig4}(b).

\section{Results for $t$-$J$ model}\label{sec:A2}

\begin{figure}[thbp]
\includegraphics[width=0.45\textwidth]{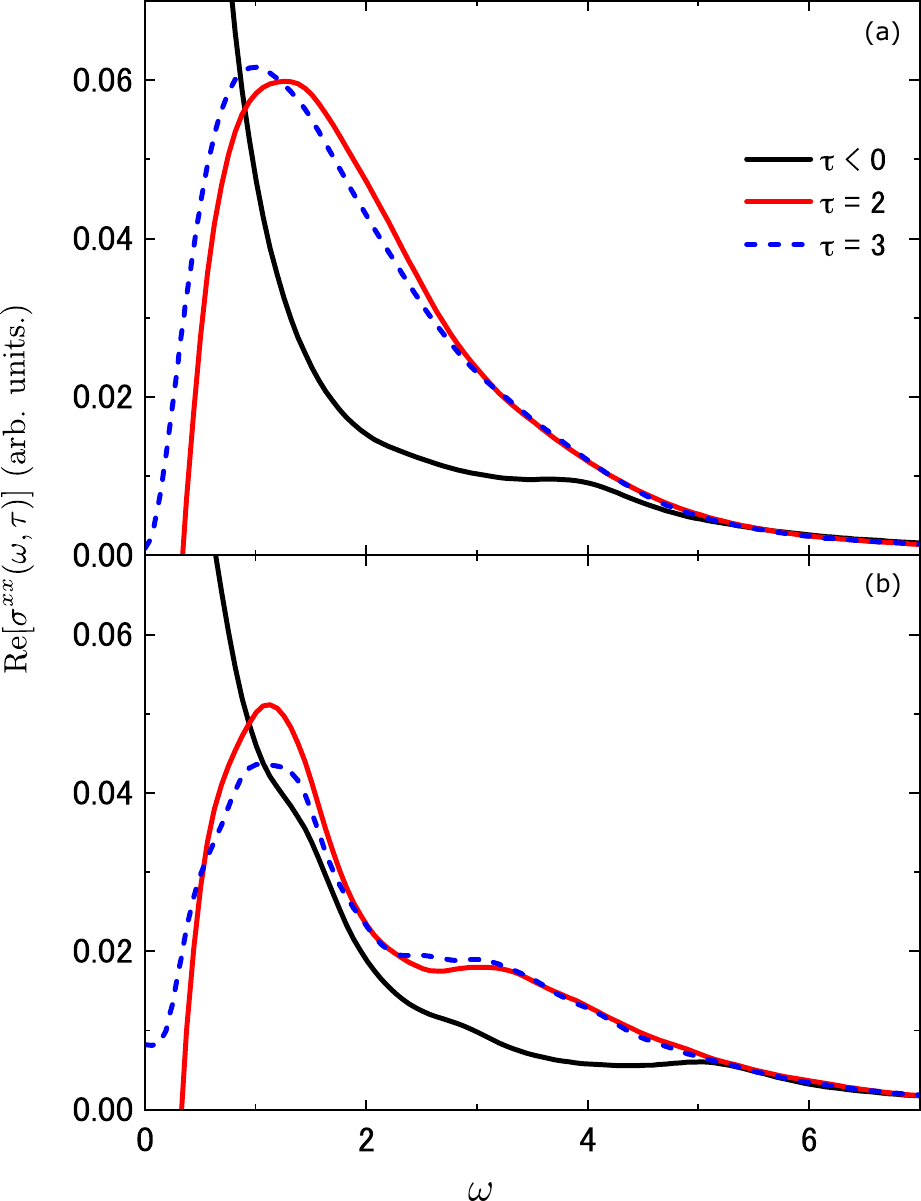}
\caption{Re$[\sigma^{xx}(\omega,\tau)]$ (parallel to a pump pulse) of the $t$-$J$ model with a hole density of $N/L=1/8$ on a $L=32\times 2$ cluster for (a) $J=1$ and (b) $J=2$. 
The black lines represent the results obtained before pumping ($\tau<0$).
The red and blue lines show the results obtained after applying a pump pulse with $A_{0}=0.3\pi$ at $\tau=2$ and 3, respectively.}
\label{FigA1}
\end{figure}

In this appendix, we show that the centroid energy of absorptions in the $t$-$J$ model, excited by an ultrashort half-cycle pulse, shifts to higher energies as the spin-exchange interaction $J$ increases. 
The Hamiltonian of the $t$-$J$ model without a vector potential $\bm{A}(t)$ is given by 
\begin{align}
\mathcal{H}_{t\text{-}J} =-t_{\rm h}\sum_{\langle i,j\rangle,\sigma} (\tilde c_{i,\sigma}^{\dag} \tilde c_{j,\sigma}+\text{H.c.})
+ J\sum_{\langle i,j\rangle} \left( \tilde{\bm{S}}_{i}\cdot \tilde{\bm{S}}_{j} -\frac{1}{4}\tilde n_{i} \tilde n_{j} \right),
\end{align}
where $\tilde c_{i,\sigma} = c_{i,\sigma}(1-n_{i,\bar\sigma})$, $\tilde n_{i,\sigma}=\tilde c_{i,\sigma}^{\dag} \tilde c_{i,\sigma}$, and 
$\bar\sigma$ represents the opposite spin to $\sigma \,=(\uparrow,\downarrow)$.  
$(\tilde{\bm{S}}_{i})_a=\frac{1}{2}\sum_{s,s'}\tilde c_{i,s}^\dag\sigma_{ss'}^a \tilde c_{i,s'}$ is the $a$-component $(a=x,y,z)$ of the spin-1/2 operator at site $i$, where $\sigma_{ss'}^a$ is the $(s,s')$ element of the $a$-component of the Pauli matrix.
The time dependent vector potential $\bm{A}(t)$ is introduced similarly as in Eq.~(\ref{eq-H}). 
Figure~\ref{FigA1} shows the results of Re$[\sigma^{xx}(\omega,\tau)]$ for the $t$-$J$ model with a hole density $N/L=1/8$ on a two-leg ladder of $L=32\times 2$ sites under open boundary conditions. 
As in the case of the Hubbard model, after pumping, we observe a suppression of the Drude weight and an enhancement of absorptions at finite energies in the range $1.5< \omega <5$. 
Comparing the results for $J=1$ and 2 in Figs.~\ref{FigA1}(a) and \ref{FigA1}(b), respectively, we find that the centroid energy of the induced absorptions shifts to higher values with increasing $J$.
Therefore, the induced absorptions can be attributed to spin-exchange interactions.


\begin{thebibliography}{99}
\bibitem{notex} A half-cycle pulse introduces a small amount of photo-doped carriers. 
In $U=-4$ ($U=-8$), photo-doped doublon density $\Delta n_\text{d} = \left[\langle \psi(\tau=2)|I|  \psi(\tau=2)\rangle -\langle \psi(\tau<0)|I| \psi(\tau<0) \rangle \right]/L$ evaluated immediately after pumping are 0.01, 0.03, and $-0.01$ (0.004, 0.02, and $-0.01$) for $A_{0}=0.1\pi$, $0.4\pi$, and $0.5\pi$, respectively. Here, $I=\sum_{i}n_{i,\uparrow}n_{i,\downarrow}$.
\bibitem{notesc} We have confirmed that the long-range behavior of a superconducting pair correlation function $P(j,\tau)=\langle \psi(\tau)| \left( \Delta_{L/2}^{\dag} \Delta_{j} + \text{H.c.} \right) |\psi(\tau)\rangle$ is almost unchanged after pumping even with $A_{0}=0.5\pi$, where $\Delta_{j}=c_{j,\uparrow}c_{j,\downarrow}$ is a pair field.
\end{thebibliography}
\end{document}